\documentclass[prd,aps,twocolumn, showpacs]{revtex4}
\usepackage[dvips]{graphicx}
\begin{document}
\newcommand{\Prd}{Phys. Rev D}
\newcommand{\Prl}{Phys. Rev. Lett.}
\newcommand{\Pl}{Phys. Lett.}
\newcommand{\Cqg}{Class. Quantum Grav.}
\newcommand{\Sch}{Schwarzschild$\;$}
\newcommand{\ochi}{\overline\chi}


\title{Analog black holes in flowing dielectrics}

\author {M. Novello,
S. Perez Bergliaffa, and J. Salim }
\affiliation{Centro Brasileiro de Pesquisas F\'{\i}sicas, \\
Rua Dr.\ Xavier Sigaud 150, Urca 22290-180 Rio de Janeiro, RJ -- Brazil}
\author{V. A. De Lorenci and R. Klippert}
\affiliation{Instituto de Ci\^encias, Universidade Federal de Itajub\'a,\\
Av. BPS 1303 Pinheirinho, 37500-903, Itajub\'a, MG - Brazil}

\begin{abstract}
\hfill{\small\bf Abstract}\hfill\smallskip
\par
We show that a flowing dielectric medium with a linear response
to an external electric field can be used to generate an analog geometry
that has many of the formal properties of a \Sch black hole for light rays, in spite of
birefringence. The surface gravity of this analog black hole has a contribution that depends
only on the dielectric properties of the fluid (in addition to
the usual term dependent on the acceleration). This
term may be give a hint to a new mechanism to increase the temperature
of Hawking radiation.
\end{abstract}
\maketitle
\newcommand{\beq}{\begin{equation}}
\newcommand{\eeq}{\end{equation}}
\newcommand{\vare}{\varepsilon}

\section{Introduction}

In recent years,
there has been a growing interest in models that mimic
in the laboratory some features of gravitation
 \cite{analog}. The actual realization of these models relies on systems that
are very different in nature:
ordinary nonviscous fluids, superfluids, flowing and non-flowing
dielectrics, non-linear electromagnetism in vacuum, and
Bose-Einstein condensates (see \cite{v1} for a complete list of references).
The basic feature shared by these systems is that the
behaviour of the fluctuations around a background solution
is governed by an ``effective metric''.
More precisely, the particles associated to the perturbations do not follow geodesics
of the
background spacetime but of a Lorentzian geometry described by the effective metric,
which depends on the background solution.
It is important
to notice that
only some kinematical aspects of general relativity can be imitated by this method
\cite{visserprl}, but
not its dynamical features (see however
\cite{v1,v2}).

This analogy has permitted the simulation of several configurations of the gravitational
field, such as wormholes and closed space-like curves for photons \cite{biwh,ctc}.
Particular attention has been paid to analog black holes, because
these would emit Hawking radiation, as was shown first by Unruh in the case of dumb black holes
\cite{unruh}.
It is the prospect of observing this radiation (thus testing the hypothesis that the thermal
emission is independent of the physics at arbitrarily short wavelengths \cite{unruh})
that motivates the quest for a realization of
analog black holes in the laboratory.
Let us emphasize that
the actual observation of the radiation is
a difficult task from the point of view of the experiment, if only
because of the extremely low temperatures
involved. In the case of a quasi one-dimensional flow
of a Bose-Einstein condensate for instance,
the temperature of the radiation would be around 70 nK, which is comparable but lower than
the temperature needed form the condensate \cite{matto}.

In this paper we would like to explore the possibilities offered
in the construction of analog black holes for photons
by nonlinear flowing dielectrics
\footnote{The generation of a black hole for photons in a nonlinear static medium
was considered previously in \cite{n1}.}.
Let us remark that, to our knowledge, all the articles
that analyze different aspects of
black holes in dielectric fluids \cite{dbh} published up to date
are
devoted to the case of a constant permittivity tensor.
Here instead we would like to present
a new static and spherically symmetric
analog black hole, generated by
a {\em flowing} isotropic dielectric that depends on an applied electric field.
As we shall see, the radius of
the horizon and the temperature of this black hole depend on three
parameters (the zeroth order permittivity, the charge that
generates the external field, and the linear susceptibility) instead of depending only
on the zeroth order permittivity. Another feature of this black hole is that
there is a new term in the surface gravity
(and hence in the temperature of Hawking radiation), in addition to the usual term
proportional to the acceleration of the fluid. This new term depends exclusively
on the dielectric properties of the fluid, and it points to a new mechanism
to get Hawking radiation with temperature higher than that reported up to date.

We shall begin with the examination of the issue of photon propagation
in a nonlinear medium in the next Section. We anticipate that photons experience
birefringence, i.e. there are two effective metrics associated to
this medium. A photon ``sees'' one of these metrics or the other
depending on its polarization. Although it may be argued that
birefringence spoils the whole effective geometry idea, we shall
show in Section \ref{analogbh} that {\em photons see one and the
same black hole, independently of their polarization}. In Section
\ref{example} we shall analyze in detail the case of a medium with
a linear dependence in the external field. Some features of the
motion of photons in this black hole
are also exhibited
in this section for the special case of
a constant velocity flow, and it will be shown that there is a simple
relation between the two effective metrics.
In Section \ref{new} we calculate the effective surface gravity of these black holes for an
arbitrary velocity profile, and display the appearance of the new term.
We close in Section
\ref{conc} with a discussion of the results.


\section{Propagating modes}

In this section we shall study the propagation of photons with different polarizations in a
nonlinear medium (see Ref.\cite{Souza} for details and notation).
Let us define first the antisymmetric tensors $F_{\mu\nu}$ and
$P_{\mu\nu}$ representing the electromagnetic field.
They can be expressed in terms of the strengths ($E$, $H$) and
the excitations ($D$, $B$) of the electric and magnetic fields as
\begin{eqnarray*}
F_{\mu\nu} &=& v_{\mu}E_{\nu} - v_{\nu}E_{\mu}
- \eta_{\mu\nu}{}^{\alpha\beta}v_{\alpha}B_{\beta},
\label{1}
\\
P_{\mu\nu} &=& v_{\mu}D_{\nu} - v_{\nu}D_{\mu}
- \eta_{\mu\nu}{}^{\alpha\beta}v_{\alpha}H_{\beta} .
\label{2}
\end{eqnarray*}
where $v_\mu$ represents the 4-velocity of an
arbitrary observer (which we will take later as comoving
with the fluid). The Levi-Civita tensor
introduced above is defined in such way that $\eta^{0123} = +1$ in Cartesian coordinates.
Since the electric and magnetic fields are spacelike vectors,
the notation
\begin{math}
E^{\alpha}E_{\alpha}\equiv -E^{2},
H^{\alpha}H_{\alpha}\equiv -H^{2}
\end{math}
will be used.
We will consider here media with properties determined only by the tensors
$\epsilon_{\alpha\beta}$ and $\mu_{\alpha\beta}$ ({\em i.e.} media with
null magneto-electric tensor), which relate the electromagnetic excitations
to the field strengths by the constitutive laws,
\beq
D_{\alpha} =  \epsilon_{\alpha}{}^{\beta}(E,H)E_{\beta},\;\;\;\;\;\;\;\;
B_{\alpha} = \mu_{\alpha}{}^{\beta}(E,H)H_{\beta}.
\label{consteq}
\eeq
In order to get the effective metric, we shall use Hadamard's method \cite{hadamard}.
The discontinuity of a function $J$ through a surface $\Sigma$ will be represented by
the symbol $[J]_\Sigma$, and its definition is
$$
[J]_\Sigma \equiv \lim_{\delta\rightarrow 0^+} \left( \left. J\right|_{\Sigma +\delta}
- \left. J \right|_{\Sigma - \delta}\right) .
$$
By taking the discontinuity of the field equations
$^*F^{\mu\nu}{}_{;\nu}=0$ and $P^{\mu\nu}{}_{;\nu}=0$, and assuming that
\beq
\epsilon^{\mu\beta} = \epsilon(E)(\eta^{\mu\beta}-v^{\mu}v^{\beta}),
\label{epsilon}
\end{equation}
and
\beq
\mu^{\mu\beta} = \mu_0(\eta^{\mu\beta}-v^{\mu}v^{\beta}),
\label{mu0}
\end{equation}
with $\mu_0 = $ const., we get
\beq
\epsilon (k.e) - \frac{\epsilon'}{E} (E.e)(k.E) = 0
\label{ke}
\eeq
\beq
\mu_0 (k.h)=0
\eeq
\beq
\epsilon (k.v) e^\mu - \frac{\epsilon'}{E} E^\alpha e_\alpha (k.v) E^\mu +
\eta^{\mu\nu\alpha\beta} k_\nu v_\alpha h_\beta = 0
\label{eom3}
\eeq
\beq
\mu_0 (k.v) h^\mu - \eta^{\mu\nu\alpha\beta} k_\nu v_\alpha e_\beta = 0
\label{eom4}
\eeq
where $k^\mu$ is the wave propagation vector, $\epsilon '$ is the derivative of $\epsilon$ w.r.t.
$E$, and
$$
[E_{\mu ,\lambda}] = e_\mu\;k_\lambda ,\;\;\;\;\;\;\;\; [H_{\mu ,\lambda}]= h_\mu\;k_\lambda .
$$
Note in particular that Eqn.(\ref{ke}) shows that the vectors $k^\mu$ and $e^\mu$ are not
always orthogonal, as would be the case if $\epsilon '$ was zero.

Substituting Eqn.(\ref{eom4}) in (\ref{eom3}), we get
\beq
Z^{\mu\beta}e_{\beta} = 0
\label{53}
\eeq
where the matrix $Z$ is given by
\begin{eqnarray}
Z^{\mu\beta} & = & \left[ k^2 + (k.v)^2 (\mu_0 \epsilon -1) \right]
\eta^{\mu\beta} \nonumber \\
 & & -\mu_0 \frac{\epsilon'}{E} (k.v)^2 E^\mu E^\beta
 + (v.k) (v^\mu k^\beta + k^\mu v^\beta )\nonumber \\
& &  - \left[ \epsilon \mu_0 (k.v) +k^2 \right]\;
v^\mu v^\beta - k^\mu k^\beta  .
\label{54}
\end{eqnarray}

Non-trivial solutions of Eqn.(\ref{53}) can be found only for
cases in which $\det\left| Z^{\mu\beta} \right| = 0$ (
this condition is a generalization of the well-known Fresnel equation
\cite{Landau}).

Eqn.(\ref{53}) can be solved by expanding
$e_{\nu}$ as a linear combination of the four linearly independent
vectors $v_{\nu}$, $E_{\nu}$, $k_{\nu}$ and
$\eta_{\alpha\beta\mu\nu}v^\alpha E^\beta k^\mu$ (the particular case in which the
vectors  $v_{\nu}$, $E_{\nu}$ and $k_{\nu}$ are coplanar will be examined below).
That is,
\begin{equation}
e_{\nu}=\alpha E_{\nu}
+\beta\eta_{\alpha\lambda\mu\nu}v^\alpha E^\lambda k^\mu
+\gamma k_{\nu}+\delta v_{\nu}.
\label{e-mu}
\end{equation}
Notice that taking the discontinuity of $E^\mu_{,\lambda}$ we can show that
$(e.v) = 0$. This restriction imposes a relation between the coefficients of
Eqn.(\ref{e-mu}):
$$
\delta = -\gamma (k.v)
$$
With the expression given in Eqn.(\ref{e-mu}), Eqn.~(\ref{53}) reads
\begin{eqnarray}
\alpha\left[
k^2 -\left(1 - \mu_0\;(\epsilon \; E)'
\right)(k.v)^2 \right] -\gamma\left[ \mu_0 (k.v)^2\frac{1}{E}
\;\epsilon '
E^{\alpha}k_{\alpha} \right] &=& 0 \nonumber
\\
\alpha E^{\mu}k_{\mu}+\gamma(1-\mu_0\epsilon)(k.v)^2+\delta(k.v) = 0 & & \nonumber
\\
\alpha(k.v)E^{\mu}k_{\mu}+\gamma(k.v)k^2+
\delta\left[k^2+\mu_0\epsilon\;(k.v)^2 \right]= 0 & & \nonumber
\\
\beta
\left[k^2-(1-\mu_0\epsilon)(k.v)^2\right] = 0 & &  \nonumber .
\end{eqnarray}
The solution of this system results
in the following dispersion relations:
\begin{eqnarray}
k_-^2 &=& (k.v)^2\left[1 - \mu_0(\epsilon\;  E)'
\right]
+ \frac{1}{\epsilon E}\;\epsilon'\;
E^{\alpha}E^{\beta}k_{\alpha}k_{\beta},
\label{g-5}\\
k_+^2 &=& [1-\mu_0\epsilon(E)](k.v)^2.
\label{g-6}
\end{eqnarray}
They correspond to the propagation modes
\begin{eqnarray}
e^-_\nu&=&\rho^- \left\{\mu_0\;\epsilon(k.v)^2 E_\nu
+E^\alpha k_\alpha[k_\nu-(k.v)v_\nu]\right\} ,
\label{e-}\\
e^+_\nu&=&\rho^+ \;\eta_{\alpha\lambda\mu\nu}v^\alpha E^\lambda k^\mu,
\label{e+}
\end{eqnarray}
where $\rho^-$ and $\rho^+$ are arbitrary constants.
The labels ``$+$'' and ``$-$'' refer to the ordinary and extraordinary rays,
respectively.
Eqns. ({\ref{g-5}) and ({\ref{g-6}) govern the propagation of photons in the medium characterized
by $\mu = \mu_0 = $const., and $\epsilon =
\epsilon (E)$. They can be
rewritten as $g_{\pm}^{\mu\nu}k_{\mu}k_{\nu}=0$,
where we have defined the effective geometries
\begin{eqnarray}
g_{(-)}^{\mu\nu} &=& \eta^{\mu\nu} -
\left[1 - \mu_0\; (\epsilon\; E)'\right]v^{\mu}v^{\nu}-\frac{1}{\epsilon E}
\;\epsilon '\;E^{\mu}E^{\nu},
\label{h-2}\\
g_{(+)}^{\mu\nu} &=& \eta^{\mu\nu}-[1-\mu_0\;\epsilon]v^\mu v^\nu.
\label{glmetric}
\end{eqnarray}
The metric given by Eqn.(\ref{h-2}) was derived in \cite{ns1}, while
the second metric very much resembles the metric derived by Gordon \cite{gordon}. The
difference is that in the case under consideration,
$\epsilon$ is a function of the modulus of the external electric field,
while Gordon worked with a constant
permeability.

Let us discuss now a particular instance in which the vectors used as a basis in
Eqn.(\ref{e-mu}) are not linearly independent. If we assume that
\beq
E^\mu = a k^\mu + b v^\mu ,
\label{pc}
\eeq
then, by Eqn.(\ref{e-mu}), the vectors $e^\mu$, $k^\mu$, and $v^\mu$
are coplanar. In this case, the basis chosen in Eqn.(\ref{e-mu}) is not appropriate.
Notice however that if we assume that $e^\mu$ is a combination of
vectors that are perpendicular to $k^\mu$, so that $(e.k)=0$, Eqn.(\ref{ke})
implies that $(E.e)=0$. The converse is also true: if $(E.e) =0$, then, from Eqn.(\ref{pc}),
$(k.e)=0$. For this particular case, in which $e^\mu$ is perpendicular to
$v^\mu$, $k^\mu$ (and consequently to $E^\mu$), Eqns.(\ref{53}) and (\ref{54}) imply that
$$
\left[ k^2 + (k.v)^2 (\mu_0 \epsilon -1) \right]\; e^\mu =0
$$
We see then that in the case in which $E_\mu = a k_\mu +
b v_\mu$, Fresnel's equation determines that the polarization of the photons is perpendicular
to the direction of propagation and to the velocity of the fluid. Moreover, the motion of
these photons is governed by the metric $g_{+}^{\mu\nu}$. For instance, if the electric field,
the velocity of the fluid, and the direction of propagation are all radial,
then the polarization
is in the plane perpendicular to the propagation,
and the two polarization modes feel the same geometry.

\section{The Analog Black Hole}
\label{analogbh}

We shall show in this section that the system described
by the effective
metrics given by Eqns.(\ref{h-2})-(\ref{glmetric}) can be used to produce
an analog black hole.
It will be convenient to rewrite at this point the inverse of the
effective metric given by Eqn.(\ref{h-2}) using a
different notation:
\beq
g_{\mu\nu}^{(-)} = \eta_{\mu\nu} - \frac{v_\mu v_\nu}{c^2} (1 - f) +
\frac{\xi}{1+\xi}  l_\mu l_\nu ,
\label{nsmetric}
\eeq
where we have defined the quantities
$$f \equiv \frac{1}{c^2\mu_0\epsilon (1+\xi)},\;\;\;\;\;\;\xi \equiv
 \frac{\epsilon ' E}{\epsilon},
\;\;\;\;\;\;l_\mu \equiv \frac{E_\mu}{E}.
$$
Note that
$\epsilon = \epsilon (E)$. We have introduced here the velocity of light $c$, which was
set to 1 before.

Taking a Minkowskian background in spherical coordinates, and
\beq
v_\mu = (v_0 , v_1 , 0 , 0 ), \;\;\;\;\;\;\; E_\mu =
(E_0,E_1, 0 , 0 ),
\label{vE}
\eeq
we get for the effective metric described by Eqn.(\ref{nsmetric}),
\beq
g_{00}^{(-)} = 1 - \frac{v_0^2}{c^2}\; (1 - f) +
\frac{\xi}{1+\xi} \; l_0^2 ,
\eeq
\beq g_{11}^{(-)} = -1 -\frac{v_1^2}{c^2}\; (1 - f ) +
\frac{\xi}{1+\xi} \; l_1^2 ,
\eeq
\beq
g_{01}^{(-)} = -\frac{v_0 v_1}{c^2} \;(1 - f) +
\frac{\xi}{1+\xi}\; l_0\; l_1 ,
\eeq
and $g_{22}^{(-)}$ and $g_{33}^{(-)}$ as in
Minkowski spacetime.
The vectors $v_\mu$ and $l_\mu$ satisfy the constraints
\beq
v_0^2 - v_1^2 = c^2,
\label{c4}
\eeq
\beq
l_0^2 - l_1^2 = -1,
\label{c3}
\eeq
\beq
v_0 l_0 - v_1 l_1 = 0.
\label{c2}
\eeq
This system of equations can be solved in terms of  $v_1$, and the result is
\beq
v_0^2 = c^2 + v_1^2 ,
\label{v0}
\eeq
\beq
l_0^2 = \frac{v_1^2}{c^2} ,\;\;\;\;\;\;\;\;\;\;\;\; l_1^2 = \frac{c^2+v_1^2}{c^2}.
\label{l0}
\eeq

Now we can rewrite the metric in terms of $\beta\equiv v_1/c$ \footnote{Notice that
this definition of $\beta$ coincides with the usual one for small $v_1$.}.
The explicit expression for the metric coefficients is:
\beq
g_{00}^{(-)} = \frac{1 - \beta^2(c^2\mu_0 \epsilon -1)}{c^2\mu_0 (\epsilon + \epsilon
' E)}
\label{g00},
\eeq
\beq
g_{01}^{(-)} = \beta  \sqrt{1+\beta^2}\; \frac{1-c^2\mu_0\epsilon}{c^2\mu_0
(\epsilon +\epsilon ' E)} ,
\label{g01}
\eeq
\beq
g_{11}^{(-)} = \frac{\beta^2 -c^2\mu_0\epsilon (1+\beta^2)}
{c^2\mu_0 (\epsilon + \epsilon ' E)} .
\label{g11}
\eeq
From Eqn.(\ref{g00}) it is easily seen that, depending on the function
$\epsilon(E)$, this metric has a horizon at $r=r_h$, given
by the condition $g_{00}(r_h) = 0$ or, equivalently,
\beq
\left. \left( c^2\mu_0 \epsilon - \frac{1}{\beta^2}\right) \;\right|_{r_h} = 1 .
\label{rh}
\eeq

The metric given above resembles the form of
Schwarzschild's solution in Painlev\'e-Gullstrand coordinates
\cite{pain,gull}:
\beq
ds^2 = \left(1 - \frac{2GM}{r}\right) dt^2 \pm 2\sqrt{\frac{2GM}{r}}\;dr\;dt - dr^2 - r^2 d\Omega^2 .
\label{pain}
\eeq
With the coordinate transformation
\beq
dt_{P} = dt_S \mp \frac{\sqrt{2GM/r}}{1-\frac{2GM}{r}} \;dr ,
\label{trafo}
\eeq
the line element given in Eqn.(\ref{pain}) can be written in Schwarzschild's coordinates. The
``$+$'' sign covers the future horizon and the black hole singularity.

The effective metric given by Eqns.(\ref{g00})-(\ref{g11}) looks like the metric in
Eqn.(\ref{pain}) \footnote{It should be kept in mind that
for the metric in Eqns.(\ref{g00})-(\ref{g11}) to
have $g_{11}=-1$ as in Painlev\'e-Gullstrand, a simple conformal transformation
is needed. Consequently, the metric of the "Schwarzchild black hole" presented below is
actually conformal to the Schwarzchild metric.}. In fact, it can be
written in
Schwarzschild's coordinates, with the coordinate change
\beq
dt_{PG}  = dt_S - \frac{g_{01}(r)}{g_{00}(r)} dr .
\label{tr}
\eeq
Using this transformation with the metric coefficients given in Eqns.(\ref{g00}) and (\ref{g01}),
we get the expression of $g_{11}^{(-)}$ in \Sch coordinates:
\beq
g_{11}^{(-)} = -\frac{\epsilon(E)}{(1-\beta^2[c^2\mu_0\epsilon(E) -1])(\epsilon(E) + \epsilon(E) 'E)}.
\eeq
Note that $g_{01}^{(-)}$ is
zero in the new coordinate system, while
$g_{00}^{(-)}$ is still given by Eqn.(\ref{g00}). Consequently,
the position of the horizon does not change, and is still given by Eqn.(\ref{rh}).

Working in Painlev\'e-Gullstrand coordinates, we have shown that
the metric for the ``$-$'' polarization describes a Schwarzschild black hole
if Eqn.(\ref{rh}) has a solution. Afterwards we have
rewritten the ``$-$'' metric in the more familiar \Sch coordinates.
Let us consider now photons with the other polarization.
They ``see'' the metric given by Eqn.(\ref{glmetric}), whose inverse is given by:
\beq
g_{\mu\nu}^{(+)} = \eta_{\mu\nu} - \frac{v_\mu}{c}\frac{v_\nu}{c}
 \left( 1 - \frac{1}{c^2\mu_0\epsilon(E)} \right).
\eeq
Using this equation and Eqns.(\ref{v0}) and (\ref{l0}) it is straightforward to show that
\beq
g_{00}^{(+)} = 1-\left( 1+\beta^2\right)\left( 1- \frac{1}{c^2\mu_0\epsilon(E)}\right),
\label{g002}
\eeq
\beq
g_{01}^{(+)}=-\beta\sqrt{1+\beta^2}\left( 1- \frac{1}{c^2\mu_0\epsilon(E)}\right),
\eeq
\beq
g_{11}^{(+)} = -1 - \beta^2 \left( 1- \frac{1}{c^2\mu_0\epsilon(E)}\right).
\eeq
This metric also corresponds to a \Sch black hole, for some $\epsilon(E)$ and $\beta$.
Comparing Eqns.(\ref{g00}) and (\ref{g002})
we see that the horizon of both analog black holes is located at $r_h$, given by
Eqn.(\ref{rh}).

By means of the coordinate change defined by Eqn.(\ref{tr}),
we can write
this metric in
Schwarzschild's coordinates. The relevant coefficients are given by
\beq
g_{00}^{(+)} = \frac{1+\beta^2(1-c^2\mu_0\epsilon(E))}{c^2\mu_0\epsilon(E)},
\label{g00m}
\eeq
\beq
g_{11}^{(+)} = -\frac{1}{1+ \beta^2 (1-c^2\mu_0 \epsilon(E))}.
\eeq

It is important to stress then
that {\em the horizon is located at $r_h$ given by
Eqn.(\ref{rh}) for photons with any polarization}.
Moreover, the motion of the photons in both geometries
will be qualitatively the same, as we shall show
below.

\section{An example}
\label{example}

We have not specified up to now the functions $\epsilon(E)$ and
$E(r)$ that determine the dependence of the coefficients of the
effective metrics with the coordinate $r$. From now on we assume a
linear $\epsilon(E)$
\footnote{This type of behaviour is
exhibited for instance by electrorheological fluids. See for instance
W. Wen, S. Men, and K. Lu, Phys. Rev. E {\bf 55}, 3015 (1997).
}.},
\beq \epsilon(E) = \epsilon_0 (\overline\chi +
\chi^{(2)} E(r)) ,
\label{eps}
\eeq
with
$\overline\chi =
1+\chi^{(1)}$. The nontrivial Maxwell's equation then reads
\beq \left(
\sqrt{-\gamma}\;\epsilon (r) F^{01}\right)_{,1} = 0,
\label{max}
\eeq where $\gamma$ is the determinant of the flat
background metric. Taking into account that \beq (F^{01})^2 =
\frac{E^2}{c^2}, \eeq we get as a solution of Eqn.(\ref{max}) for
a point source in a flat background in spherical coordinates
\beq
F^{01} = \frac{-\overline\chi \pm \sqrt{\ochi^2 + 4\chi^{(2)}
Q/\epsilon_0 r^2}}{2c\chi^{(2)}} .
\eeq
Let us consider a
particular combination of parameters: $\chi^{(2)} >0$, $Q > 0$ and
the ``$+$'' sign in front of the square root in $F^{01}$, in such
a way that $E>0$ for all $r$. To get more manageable expressions
for the metric, it is convenient to define the function $\sigma
(r)$: \beq E(r) \equiv \frac{\ochi}{2\chi^{(2)}}\; \sigma (r) \eeq
where \beq \sigma (r) = -1 + \frac{1}{r} \sqrt{r^2 + q}
\label{sigma}
\eeq
and
\beq
q = \frac{4\chi^{(2)} Q}{\epsilon_0\ochi ^2 }.
\label{q}
\eeq
In terms of $\sigma$, the metrics take the form
\begin{eqnarray}
ds_{(-)}^2 & = & \frac{2-\beta^2\;[\;\ochi\; (\sigma (r)+2) -2]}{2\;\ochi\;
 (1+\sigma(r))}\; d\tau ^2 -
\\ \nonumber
&  & \frac{2+\sigma (r)}
{[2-\beta^2\;(\ochi\; (\sigma (r) +2)-2)]\;(1+\sigma (r))}
\;dr^2 - r^2 d\Omega^2,
\label{nsm}
\end{eqnarray}
\begin{eqnarray}
ds_{(+)}^2 & = & \frac{2-\beta^2\;[ \;\ochi\;(\sigma (r) +2)-2]}{\ochi\; (2+\sigma (r))}\; d\tau^2
 - \\ \nonumber
 & & \frac{2}{2+\beta^2\;[2-\ochi\; (\sigma (r)+2)]}\; dr^2 - r^2 d\Omega^2.
\label{gm}
\end{eqnarray}

Notice that the $(t,r)$ sectors of these metrics
are related by the following expression:
\beq
ds^2_{(+)} = \Phi (r)\; ds^2_{(-)}
\label{confinv}
\eeq
where the conformal factor $\Phi$ is given by:
$$
\Phi = 2\;\frac{1+\sigma (r)}{2+\sigma (r)}
$$

We shall study next some features of the effective black hole metrics.
We would like to remark that up to this point, the velocity of the fluid
$v_1$ is completely
arbitrary; it can even be a function of the coordinate $r$. We shall assume in the following
that $v_1$ is a constant. This assumption, which will be lifted in
Sect.\ref{new}, may seem rather restrictive but it helps to
display the main features of the effective metrics in an easy way.

To study the motion of the photons in these geometries, we can use the technique of
the effective potential.
Standard manipulations (see for instance \cite{wald}) show that in the case of a
static and spherically symmetric metric, the effective potential is given by
\beq
V(r) = \varepsilon^2 \left( 1 + \frac{1}{g_{00}(r)\;g_{11}(r)}\right) - \frac{L^2}{r^2
g_{11}(r)}
\label{effpot}
\eeq
where $\varepsilon$ is the energy and $L$ the angular momentum of the photon.

In terms of $\sigma (r)$, and of the impact parameter $b^2 = L^2/\varepsilon^2$, the
"small" effective potential $v(r)\equiv V(r)/\varepsilon^2$ for the metric Eqn.(\ref{nsm})
 in \Sch coordinates can be written as follows:
\beq
v^{(-)}(r)  =  1-\frac{2(1+\sigma (r))^2}{2+\sigma (r)}
 -\frac{b^2}{r^2} \frac{(2-\beta^2\sigma (r))(1+\sigma (r))}{2+\sigma (r)}
\eeq
A short calculation shows that $v^{(-)}$
is a monotonically decreasing function of $\beta$. Consequently,
we shall choose a convenient value
of it, for the sake of illustrating the features of the effective potential.
Figures (\ref{effpot1}) and (\ref{effpot2}) show the plots
of the potential for the $(-)$ metric
for several values of the relevant parameters.
\begin{figure}[h]
\begin{center}
\includegraphics[angle=-90,width=0.5\textwidth]{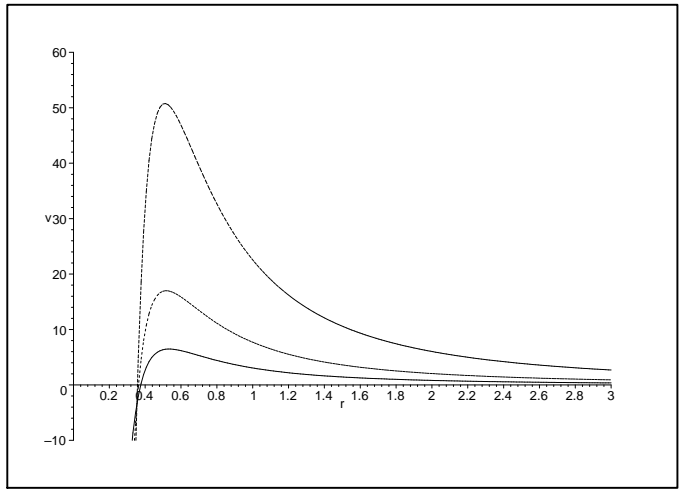}
\caption{Plot of the effective potential $v^{(-)}(r)$ for $q=1$,
$b=1,3,5$ (starting from the lowest curve), and $\beta=0.5$.}
\label{effpot1}
\end{center}
\end{figure}
\begin{figure}[h]
\begin{center}
\includegraphics[angle=-90,width=0.5\textwidth]{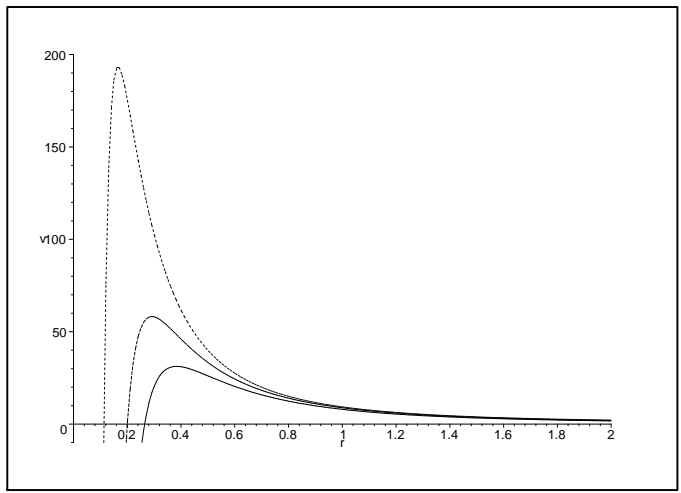}
\caption{Plot of the effective potential $v^{(-)} (r)$
for $b=3$, and $q=1, 3, 5$ (starting from the lowest
curve), and $\beta = 0.5$.}
\label{effpot2}
\end{center}
\end{figure}

The effective potential for the Gordon-like metric
can be obtained in the same way. From Eqns.(\ref{effpot}) and (\ref{gm}) we get
\beq
v^{(+)}(r) = 1 - \frac{2+\sigma (r)}{2}+ \frac{b^2}{2r^2}\;[2-\beta^2\sigma (r)]
\label{geffpot}
\eeq
The plots in Figures (\ref{effpot3}) and (\ref{effpot4})
show the dependence of $v^{(+)}(r)$ on the different parameters.
\begin{figure}[h]
\includegraphics[angle=-90,width=0.5\textwidth]{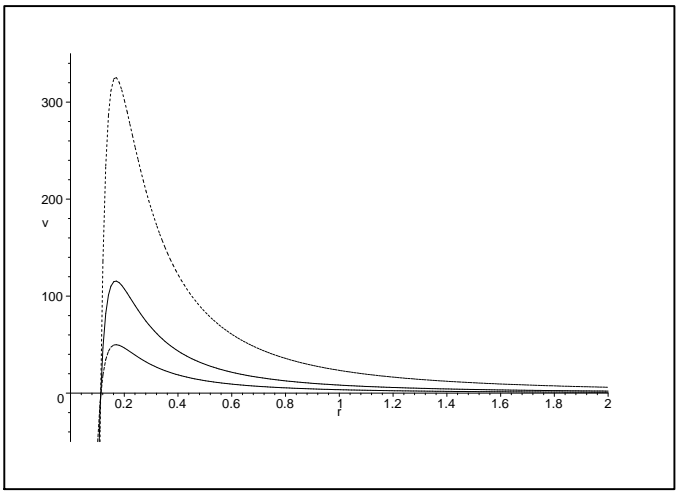}
\caption{Plot of the effective potential for the Gordon-like metric,
for $q=1$, $b=1, 3, 5$ (starting from the lowest curve), and $\beta = 0.5$.}
\label{effpot3}
\end{figure}

\begin{figure}[h]
\includegraphics[angle=-90,width=0.5\textwidth]{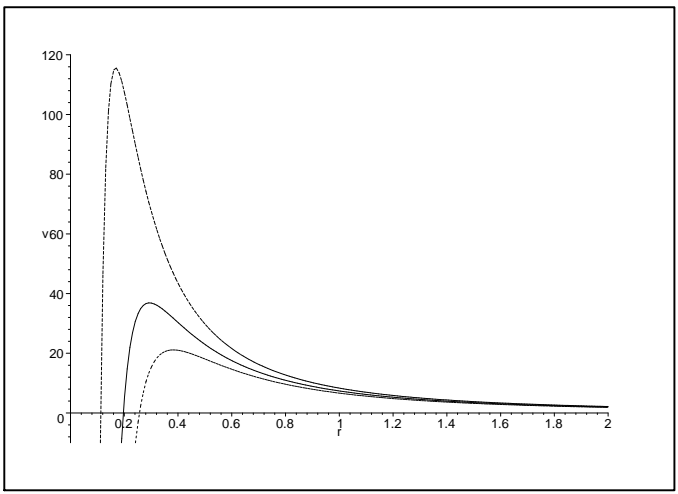}
\caption{Plot of the effective potential for the Gordon-like metric,
for $b=3$, $q=1, 3, 5$.}
\label{effpot4}
\end{figure}

We see from these plots that, in the case of a constant flux velocity,
the shape of the effective potential for both metrics qualitatively agrees with
that for photons moving on the geometry of a \Sch black hole
(see for instance Ref.\cite{wald}, pag. 143).

\section{Surface gravity and temperature}
\label{new}

Let us now go back to the more general case of $\beta = \beta(r)$,
and calculate the ``surface gravity'' of our analog black hole.
We present first the results for the constant permittivity case.
By setting
$\epsilon'(E)\equiv 0$ in the metrics Eqns.(\ref{h-2}) and (\ref{glmetric}),
we regain the example of constant
index of refraction studied for instance
in \cite{dbh}. It is easy to show from Eqn.(\ref{rh}) that the horizon of the black hole
in this case is given by
\beq
\beta^2(r_h) = \frac{1}{\bar\chi -1}.
\label{rad}
\eeq
The ``surface
gravity'' of
a spherically symmetric analog black hole in Schwarszchild coordinates
is given by \cite{viss}
\beq
\kappa = \frac {c^2}{ 2} \lim_{r\rightarrow r_h}\; \frac{g_{00,r}}{\sqrt{|g_{11}|
\;g_{00}}}.
\label{kappa}
\eeq
For the metrics Eqns.(\ref{h-2}) and (\ref{glmetric}) with
$\epsilon = \epsilon_0\bar\chi$ and $r_h$ given by
Eqn.(\ref{rad}), the analog surface gravity is
\beq
\kappa = - \frac{c^2}{2}\frac{1-\bar\chi}{\sqrt{\bar\chi}}\left.(\beta^2)'\right|_{r_h}.
\label{kappa1}
\eeq
In this expression we can see the influence of the dielectric properties of the fluid
(through the constant $\bar\chi$)
and also of its dynamics through the physical acceleration in the radial direction,
given by
$$
\left. a_r\right|_{r_h} = \frac{c^2}{2}\left.(\beta^2)'\right|_{r_h},
$$
for $\beta^2(r_h)\ll 1$. This acceleration
is a quantity that must
be determined solving the equations of motion of the fluid
\footnote{If we set $\beta\equiv 0$ in Eqns.(\ref{g00})-(\ref{g11}), we cease to
have a black hole (this situation was analyzed in \cite{n1}).}.

Going back the the more general case of a linear permittivity, described
by the metrics given by Eqns.(\ref{nsm}) and (\ref{gm}), and considering that
$\beta(r_h)\ll 1$,
the
radius of the horizon is \footnote{Notice that we cannot take the limit $q\rightarrow
0$ in this expression or in any expression in which this one has been used.}:
\beq
r_h^2 = \frac{q\bar\chi^2}{4} \beta^4 (r_h).
\eeq
Using the expressions given above,
the result for the surface gravity of the ``$-$'' black hole for $\beta(r_h)\ll 1$ is
\beq
\kappa^{(-)} = \left. \frac{c^2}{\beta}\left( \frac{1}{\bar\chi \sqrt q}  -
\frac 1 2 (\beta^2)'
\right)\right|_{r_h}
\label{kappa2}
\eeq
This equation
differs from the surface gravity of the case of constant permittivity
(Eqn.(\ref{kappa1}))
by the presence of a new term that does not depend on the acceleration of the fluid.
To see where this new term comes from,
we can take the derivative of the metric coefficient $g_{00}$, given by Eqn.(\ref{g00}),
with respect to the coordinate $r$
(see the definition of $\kappa$, Eqn.(\ref{kappa}). Taking into
account Eqn.(\ref{eps}), the result is
\beq
g_{00,r} = 2\;\frac{ \beta\beta' \;(\epsilon(E) + \epsilon(E)'E)\;(1-c^2\mu_0\epsilon(E))-
\frac{d\epsilon(E)}
{dr}}
{c^2\mu_0\;(\epsilon(E)+\epsilon(E)'E)^2}
\label{goor}
\eeq
In this expression, the first term is the generalization of the term corresponding to the
case $\epsilon=$ const.
(compare with Eqn.(\ref{kappa1})), which mixes the acceleration of the fluid with
its dielectric properties. On the other hand, the second term, which is the
new term
displayed in Eqn.(\ref{kappa2}),
is related only to the dielectric properties of the fluid.

It is easy to show that the these results also apply to
the black hole described by the Gordon-like
metric. This is not surprising though, because of the conformal relation
between the two metrics, given by Eqn.(\ref{confinv}) \cite{jac}.

\section{Discussion}
\label{conc}

We have shown that a flowing inhomogeneous dielectric that depends
linearly on an external electric field generates one effective
metric for each polarization state of photons. A particular configuration of the fluid (
{\em i.e.} pure radial flow) plus
an external electric field was shown to be an analog black hole, with a
radius that depends on the function $\epsilon (E)$. The features of the black hole depend
on the charge that generates the electric field, on the properties of the dielectric, and
on the velocity profile.
Note in particular
that in the absence of flux ($\beta =0$) the metrics do not describe a black hole.

The existence of two metrics reflect the birefringent properties
of the medium. Although some claims have been made that
birefringent materials spoil from the beginning the idea of an
effective metric, we have shown here that for a special
configuration, even if two metrics are present, photons with
different polarizations experience the same horizon. Moreover, as
seen from the plots of the effective potential, the motion of
these photons in the medium will depend on their polarization, but
is qualitatively the same for both types of photons.

At first sight it may seem that by choosing an appropriate material and a convenient
value of the charge we could obtain a high value of the temperature of the radiation,
given by
\footnote{The concept of temperature, and indeed that of effective geometry
is valid in this context only for photons with wavelengths long compared to the
intermolecular spacing in the fluid.
For shorter wavelengths, there would be corrections to the propagation dictated by
the effective metric. However,
results
for other systems (such as dumb black holes \cite{unruh}
and Bose-Einstein condensates)
suggest that the phenomenon of Hawking
radiation is robust ({\em i.e.} independent of this "high-energy" physics).},
\beq
T\equiv \frac{\hbar }{2\pi k_B c}\;\kappa \approx 4\times 10^{-21}\;\kappa\; {\rm Ks}^2{\rm
/m},
\label{temp}
\eeq
with $\kappa$ given in Eqn.(\ref{kappa2}).
However, the equation for the surface gravity can be rewritten as
$$
\kappa = c^2 \left.\left(\frac{\beta}{2r} - \beta'
\right)\right|_{r_h}.
\label{kappa3}
$$
We see then that, because $\beta(r_h)\ll 1$,
the new term appearing in $\kappa$ is bound to be very small, but of the same
order of magnitude of the acceleration term. To summarize,
we have shown that a new term emerges in the surface gravity which is
produced by the properties of the
material and is comparable in magnitude to the acceleration term. This result 
indicates that it may be worth to study if
some media with nonlinear dependence on an external electric field
are appropriate to generate analog black holes with Hawking radiation with a
higher temperature
than that produced up to date with other systems.

\section*{Acknowledgements}

The authors would like to thank CNPq, FAPERj, and FAPEMiG for financial support.

\end{document}